\documentclass[a4paper]{article}

\usepackage{itgspeech2021}    
\usepackage{times}            
\usepackage[english]{babel}   
\usepackage[utf8]{inputenc}
\usepackage[T1]{fontenc}      
\usepackage[sort&compress,numbers]{natbib}	
\usepackage{amsmath,amssymb}
\usepackage{graphicx}
\usepackage[colorlinks=false,pdfborder={0 0 0}]{hyperref}
\usepackage{units}

\DeclareMathAlphabet{\mathcal}{OMS}{cmsy}{m}{n} 
\usepackage{booktabs}
\usepackage{acronym}
\usepackage[tight-spacing=true]{siunitx} 
\acrodef{AER}{acoustic echo reduction}
\acrodef{RIR}{room impulse response}
\acrodef{DNN}{deep neural network}
\acrodef{DPRNN}{dual-path recurrent neural network}
\acrodef{TCN}{temporal convolution network}
\acrodef{SIR}{signal-to-interference ratio}
\acrodef{SI-SDR}{scale-invariant source-to-distortion ratio}
\acrodef{SDR}{source-to-distortion ratio}
\acrodef{TV}{time-varying}
\acrodef{TI}{time-invariant}
\acrodef{gln}[gLN]{global layer normalization}
\acrodef{ERLE}{echo return loss enhancement}
\acrodef{LSTM}{long short-term memory}

\usepackage{flushend} 
\usepackage{xurl} 

\title{Informed Source Extraction With Application to Acoustic Echo Reduction}

\author{Mohamed Elminshawi, Wolfgang Mack, Emanu\"{e}l A.~P.~Habets}

\address{International Audio Laboratories, Erlangen, Germany\\
  Email: \texttt{\{mohamed.elminshawi, wolfgang.mack, emanuel.habets\}@audiolabs-erlangen.de}}

\begin{document}

\maketitle

%

\begin{abstract}
Informed speaker extraction aims to extract a target speech signal from a mixture of sources given prior knowledge about the desired speaker. Recent deep learning-based methods leverage a speaker discriminative model that maps a reference snippet uttered by the target speaker into a \textit{single} embedding vector that encapsulates the characteristics of the target speaker. However, such modeling deliberately neglects the time-varying properties of the reference signal. In this work, we assume that a reference signal is available that is temporally correlated with the target signal. To take this correlation into account, we propose a time-varying source discriminative model that captures the temporal dynamics of the reference signal. We also show that existing methods and the proposed method can be generalized to non-speech sources as well. Experimental results demonstrate that the proposed method significantly improves the extraction performance when applied in an \acl{AER} scenario.

\end{abstract}



\acresetall

\section{Introduction}
Real-world acoustic scenes often comprise more than one audio source. Extracting a specific source while suppressing the others is crucial for many audio processing systems, such as automatic speech recognition. The state-of-the-art of this challenging problem has been significantly advanced thanks to deep learning-based approaches \cite{wang2018voicefilter, wang2018deep, vzmolikova2019speakerbeam}. Such approaches leverage prior knowledge about the target speaker, typically in the form of a reference signal or enrollment utterance, which guides the extraction system towards the desired speaker. The majority of such works have focused only on speaker extraction where the target source is always a speech signal. There are, however, very few works that considered non-speech signals as targets \cite{lee14audio, Tzinis2020}.



The authors in \cite{vzmolikova2019speakerbeam, Delcroix2020} proposed a speech extraction framework that consists of two \acp{DNN}, an auxiliary network, which acts as a speaker discriminative model, that maps the reference signal to a single embedding vector capturing the characteristics of the target speaker, and an extraction network that processes the mixture signal as well as the embedding vector and outputs an estimate of the target signal. Both networks are optimized jointly in an end-to-end fashion.



One particularly interesting scenario happens when a temporal correlation exists between the reference signal and the target signal. An example of this scenario is in \ac{AER}, where the echo signal is a filtered version of the far-end signal. Applying informed source extraction in this context aims to extract the echo given the far-end signal as a reference. One limitation of the framework in \cite{vzmolikova2019speakerbeam, Delcroix2020} is that its source discriminative model does not exploit the time-varying nature of the reference signal, which is intuitively important in this scenario. Moreover, the characteristics of the reference signal can change over time, e.g., in the presence of multiple speakers at the far-end side. Another point to consider is the fact that a successful echo reduction system should be able to eliminate non-speech echoes as well. However, the authors in \cite{vzmolikova2019speakerbeam, Delcroix2020} did not evaluate their extraction system for non-speech target signals, such as musical instruments or noise.

It is worth mentioning that time-varying discriminative embeddings have been proposed in other domains, such as audio-visual speech separation, where visual information, e.g., lip-movement, about the target speaker is provided as auxiliary features to the extraction system \cite{ephrat2018looking, wu2019time, michelsanti2021overview, Ochiai2019}.




The contribution of this paper is threefold. First, we propose a time-varying source discriminative model that takes into account the dynamics of the reference signal. Secondly, we show that current informed speaker extraction methods and the proposed method can be used for non-speech sources, such as piano or guitar. Finally, we apply the proposed method in an \ac{AER} scenario and show experimentally that it significantly improves the extraction performance compared to several baselines.




\section{Problem Formulation}
We consider an observed mixture signal $y(t)$ that comprises $C$ audio sources  $\{x_i(t)\}_{i=0}^{C-1}$ as

\begin{equation}
  y(t) = x_0(t) + \sum_{i=1}^{C-1} x_i(t),
\end{equation}

\noindent
where $t$ is the discrete-time index, and $x_0(t)$ represents the target signal emitted from the desired source. In addition, a reference waveform $a_0(t)$ is provided, which belongs to the target source but is different from $x_0(t)$. The goal of informed source extraction is to estimate $x_0(t)$ given $y(t)$ and $a_0(t)$ as inputs.

Following the time-domain masking-based approach in \cite{Delcroix2020}, the mixture and reference waveforms are first segmented into $T$ overlapping frames of length $L$, denoted by $\mathbf{y}_\kappa, \mathbf{a}_\kappa \in \mathbb{R}^{L}$, where $\kappa = 1, \ldots, T$ represents the time-frame index. The notation $\mathbf{Y} = [\mathbf{y}_{1}, \ldots, \mathbf{y}_{T}] \in \mathbb{R}^{L \times T}$ and $\mathbf{A} = [\mathbf{a}_{1}, \ldots, \mathbf{a}_{T}] \in \mathbb{R}^{L \times T}$ will be used for the matrices containing the frame-wise vectors. Moreover, two jointly optimized \acp{DNN} are employed: an auxiliary network
 and an extraction network, represented by $\mathcal{Z}$ and $\mathcal{H}$, respectively. The auxiliary network $\mathcal{Z}$ maps $\mathbf{A}$ to an embedding matrix $\mathbf{E} =[\mathbf{e}_{1}, \ldots, \mathbf{e}_{T}] \in \mathbb{R}^{N \times T}$, expressed as




\begin{equation}
  \mathbf{E} = \mathcal{Z}(\mathbf{A}),
\end{equation}


\noindent
where $\mathbf{e}_{\kappa} \in \mathbb{R}^{N}$ denotes the $N$-dimensional embedding vector for the time-frame $\kappa$. In \cite{vzmolikova2019speakerbeam, Delcroix2020}, the authors proposed to average the embeddings over the temporal dimension resulting in a single \ac{TI} embedding vector, represented by $\mathbf{\bar{e}} \in \mathbb{R}^{N}$, that encapsulates the discriminative information of the target source as

\begin{equation}
  \mathbf{\bar{e}} = \frac{1}{T} \sum_{\kappa=1}^{T} \mathbf{e}_{\kappa}.
  \label{eq:average}
\end{equation}

 The vector $\mathbf{\bar{e}}$ and $\mathbf{Y}$ are then incorporated in the extraction network $\mathcal{H}$ that estimates the target frames, denoted by $\hat{\mathbf{X}}_0 = [\hat{\mathbf{x}}_{0, 1}, \ldots, \hat{\mathbf{x}}_{0, T}] \in \mathbb{R}^{L \times T}$:


\begin{equation}
  \hat{\mathbf{X}}_0 = \mathcal{H}(\mathbf{Y}, \mathbf{\bar{e}}),
\end{equation}

\noindent
where $\hat{\mathbf{x}}_{0, \kappa} \in \mathbb{R}^{L}$ represents the estimated target signal for the time-frame $\kappa$. Finally, the overlapping time-frames are summed to reconstruct the estimated target waveform, which is represented by $\hat{x}_0(t)$.






\begin{figure*}[t]
	\centerline{\includegraphics[width=.9\textwidth]{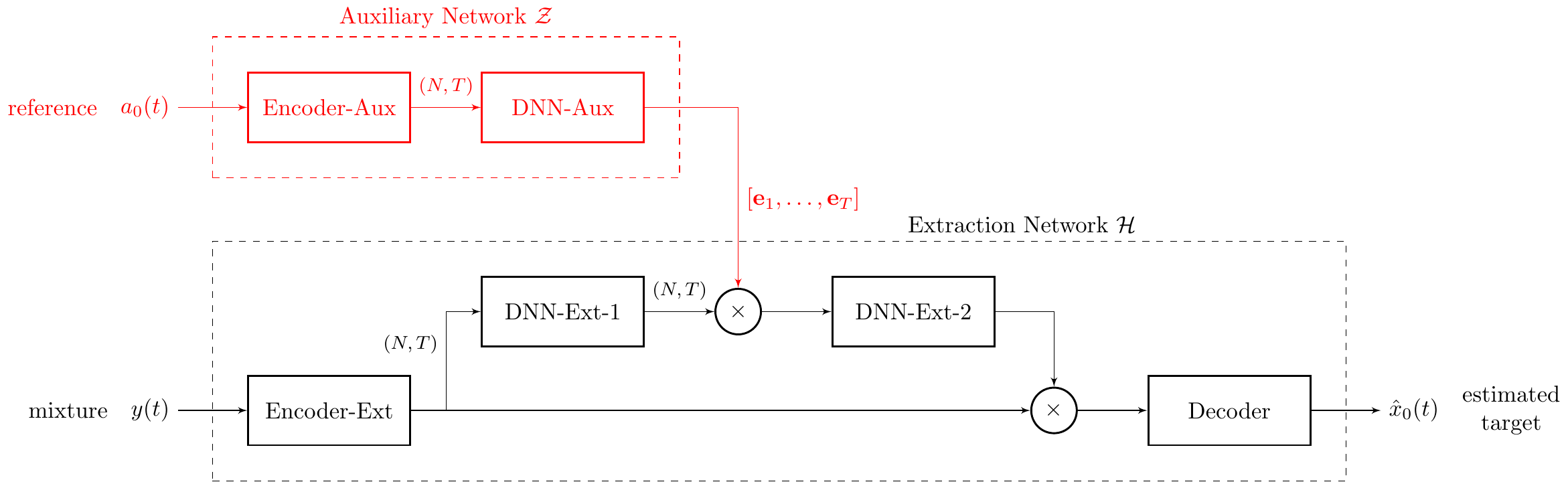}}
	\caption{The proposed \acf{TV} source discriminative embedding model for informed source extraction. The time-domain encoder-decoder framework is similar to the one in \cite{Delcroix2020}. The difference is that we incorporate the \ac{TV} embeddings $\mathbf{E} =[\mathbf{e}_{1}, \ldots, \mathbf{e}_{T}] \in \mathbb{R}^{N \times T}$ in the extraction network and not the \acf{TI} embedding $\mathbf{\bar{e}} \in \mathbb{R}^{N}$, where $\mathbf{\bar{e}} = \frac{1}{T} \sum_{\kappa=1}^{T} \mathbf{e}_{\kappa}$. The tensor dimensions are given in parentheses. Note that the blocks \{Encoder-Aux, Encoder-Ext\} do not share weights.}
  \vspace{-0.6em}
	\label{fig:pipeline}%
\end{figure*}

%
%
%
%

\section{Proposed Approach}
\subsection{Time-Varying Embeddings}
The averaging step in (\ref{eq:average}) inherently assumes that the characteristics of the reference signal do not drastically change over time. This restrictive assumption holds when the reference signal comprises only a single target source. However, for some use cases, the temporal information could be advantageous for the extraction network. To this end, we propose to include the \acf{TV} embeddings $\mathbf{E}$ in the extraction network as

\begin{equation}
  \hat{\mathbf{X}}_0 = \mathcal{H}(\mathbf{Y}, \mathbf{E}).
\end{equation}

Similar to \cite{Delcroix2020}, the learned embeddings are incorporated in the extraction network via element-wise multiplication. The detailed block diagram of the proposed method is depicted in Figure~\ref{fig:pipeline}.

\subsection{AER Formulation}
\begin{figure}[t]
  \centerline{\includegraphics[width=.9\columnwidth]{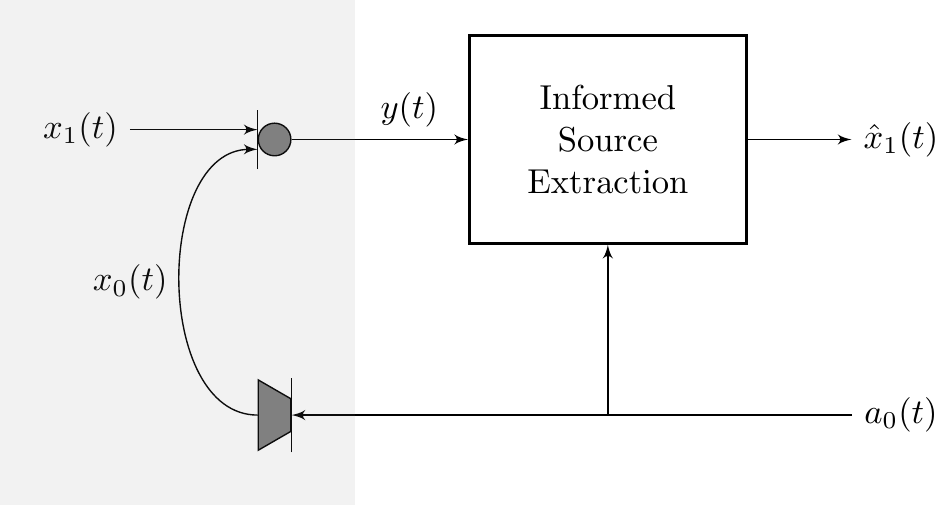}}
	\caption{The \acf{AER} scenario. Informed source extraction is used to extract the echo $x_0(t)$ from the microphone signal $y(t)$ given the far-end signal $a_0(t)$ as a reference. An estimate of the near-end signal, denoted by $\hat{x}_1(t)$, is then obtained by $\hat{x}_1(t) = y(t) - \hat{x}_0(t)$, where $\hat{x}_0(t)$ is the estimated echo.}
	\label{fig:AER}%
\end{figure}

In the rest of the paper, we apply informed source extraction in an \ac{AER} scenario. The formulation of the problem is illustrated in Figure~\ref{fig:AER}, where we assume that a microphone signal $y(t)$ consists of only two signals ($C=2$): echo $x_0(t)$ and near-end $x_1(t)$. In this context, informed source extraction is applied to extract the echo given the far-end signal as a reference $a_0(t)$. Once the echo signal is extracted, an estimate of the near-end signal, denoted by $\hat{x}_1(t)$, can be obtained by

\begin{equation}
  \hat{x}_1(t) = y(t) - \hat{x}_0(t).
\end{equation}

\subsection{Training Objective}
The most common training objective for optimizing time-domain-based source extraction systems is the \ac{SI-SDR} \cite{leroux2019, Luo2018}. One attribute of this objective function is that it does not penalize scaling mismatches between the estimated and ground-truth signals. However, in our formulation of \ac{AER}, this mismatch must be penalized since the echo is first estimated and then subtracted from the mixture to obtain the near-end signal. To this end, we propose to maximize the \ac{SDR} instead, defined as


\begin{align}
  \text{SDR} := 10 \log_{10} \frac{\sum_t{|x_0(t)|^2}}{\sum_t{|x_0(t) - \hat{x}_0(t)|^2}}.
\end{align}

\noindent
The \ac{SDR} objective function ensures that no scaling is introduced between $\hat{x}_0(t)$ and $x_0(t)$.

\section{Experimental Settings}

\subsection{Datasets}
For simulating echo and near-end signals, we generated \acp{RIR} using the image method \cite{RIRGenerator, Allen1979}. Table~\ref{tab:RIR_params} shows the parameter pool of the \ac{RIR} generator for the training, validation, and test sets.
To obtain a microphone signal, a room size, a reverberation time $T_{60}$, and two distances were randomly sampled. Given these two combinations, two \acp{RIR} were then simulated and convolved with two anechoic signals to obtain the echo and near-end signals. It should be noted that the reference signal (far-end) and the target signal (echo) were not aligned in time, i.e., no time delay compensation was applied.


We obtained the speech and non-speech signals from LibriSpeech \cite{Libri2015} and FSDnoisy18k \cite{fonseca2019learning} datasets, respectively. We restricted ourselves to the following non-speech classes: acoustic guitar, bass guitar, piano, rain, and engine. Since the FSDnoisy18k dataset does not contain a separate validation set, we reserved 30\% of the original training set for validation. We constructed four different two-source mixture subsets (in the form of target + interferer): speech + speech (SS), speech + non-speech (SN), non-speech + speech (NS), and non-speech + non-speech (NN). The duration of all signals was set to 4 seconds and a sampling frequency of 16 kHz was used. For training, validation, and testing, the mixing \ac{SIR} was sampled from -5 to 5 dB.

\begin{table}[t]
  \centering
  \resizebox{0.9\columnwidth}{!}{
\begin{tabular}{@{}lccc@{}}
\toprule
Set              & Room Size [m] & $T_{60}$ [s]  & Distance [m]   \\ \midrule
Training         & (2.0, 4.0, 2.7)  & 0.20 & 0.50 \\
                 & (6.0, 6.0, 2.7)  & 0.30 & 0.70 \\
                 & (10.0, 4.0, 2.7) & 0.40 & 0.90 \\
                 & (7.0, 3.0, 2.7)  & 0.50 & 1.10 \\
                 & (8.0, 10.0, 2.7) &      & 1.30 \\
                 &                  &      & 1.50 \\
                 &                  &      & 1.70 \\
                 &                  &      & 1.90 \\ \midrule
Validation       & (5.0, 6.0, 2.7)  & 0.23 & 0.55 \\
                 & (4.0, 3.0, 2.7)  & 0.33 & 1.05 \\
                 & (8.0, 9.0, 2.7)  & 0.43 & 1.55 \\
                 &                  & 0.53 & 2.05 \\ \midrule
Test             & (3.0, 5.0, 3.0)  & 0.25 & 0.85 \\
                 & (4.0, 6.0, 3.0)  & 0.35 & 1.35 \\
                 & (9.0, 9.0, 3.0)  & 0.45 & 1.85 \\ \bottomrule
\end{tabular}
}
\caption{Parameters of the \acf{RIR} generator for the training, validation, and test sets. The room sizes correspond to (width, length, height). Distance refers to how far the source is from the microphone.}
\vspace{-1.2em}
\label{tab:RIR_params}
\end{table}

\subsection{Network Configuration} 
In \cite{Delcroix2020}, the authors used the \ac{TCN} architecture \cite{Luo2019} for the blocks \{DNN-Aux, DNN-Ext-1, DNN-Ext-2\} shown in Figure~\ref{fig:pipeline}. We investigated employing the \ac{DPRNN} architecture since it has proven its superiority in \cite{Luo2019a}. The publicly available implementation of both architectures provided by SpeechBrain \cite{SB2021} was used in this work. We considered a causal and a non-causal version of the \ac{DPRNN} architecture. For the non-causal \ac{DPRNN} architecture, 2 \ac{DPRNN} blocks were used, and bidirectional \ac{LSTM} networks \cite{Hochreiter1997} with 128 hidden units in each direction of the intra- and inter-chunks were employed. A bottleneck size\footnote{In SpeechBrain, the bottleneck size is referred to as out\_channels.} of 64, a chunk size of 30, and \ac{gln} were used. The hyperparameters of the causal \ac{DPRNN} architecture were similar to the non-causal version, except that a uni-directional \ac{LSTM} for the inter-chunk processing and cumulative layer normalization were used instead. For the \ac{TCN} architecture, the hyperparameters were $B=64$, $H=96$, $P=3$, $X=6$, $R=2$, and \ac{gln}, following the notations in the original paper \cite{Luo2019}. ReLU non-linearity was used at the output of both architectures.

The encoders and decoder were learnable similar to \cite{Luo2018}, for which we used a window size $L$ and stride of 32 and 16 samples, respectively. The number of channels $N$ at the output of both encoders was set to 256.



\subsection{Training Setup}
\label{sec:training_setup}
For training the extraction framework, ADAM optimizer \cite{Kingma2015} was used with an initial learning rate of $10^{-3}$ and a weight-decay of $10^{-5}$. Gradients were clipped if the $\ell_{2}$ norm exceeds 5 and the batch size was set to 24. The maximum number of epochs was set to 300 epochs, where each epoch consisted of 10,000 training examples and 4,000 validation examples. The training examples were dynamically generated, i.e., not fixed for each epoch, and each example was randomly sampled from the subsets \{SS, SN, NS, NN\}. The learning rate was reduced by a factor of two if no reduction in the validation loss occurred in 10 consecutive epochs, and early stopping with a patience of 20 epochs was used.


\section{Performance Evaluation}
We compared the proposed \acf{TV} source discriminative model to the \acf{TI} model for both non-causal versions of the \ac{TCN} and \ac{DPRNN} architectures. Note that the non-causal \ac{TCN}-TI model corresponds to the baseline in \cite{Delcroix2020}. We also trained a causal version of the proposed \ac{DPRNN}-TV since the non-causal version is of limited use in a realistic \ac{AER} scenario. As an additional baseline, we implemented and trained a recent deep learning-based \ac{AER} method \cite{Zhang2018} that estimates a time-frequency mask for the near-end signal directly\footnote{Since there is no public implementation for this baseline, we implemented and trained it according to the training setup in Section~\ref{sec:training_setup}, except the learning rate is 0.0003 similar to the original paper.}. For evaluating the different methods, the \ac{SI-SDR} improvement \cite{leroux2019} is used and computed with respect to the near-end signal as it is the signal of interest in the \ac{AER} scenario. We report the average results separately for the four different subsets \{SS, SN, NS, NN\}, where each subset consists of 1,000 examples\footnote{Audio examples can be found at \url {https://www.audiolabs-erlangen.de/resources/2021-ITGspeech-Informed-Source-Extraction}}.

\begin{figure*}[t]
    \centerline{\includegraphics[width=0.9\textwidth]{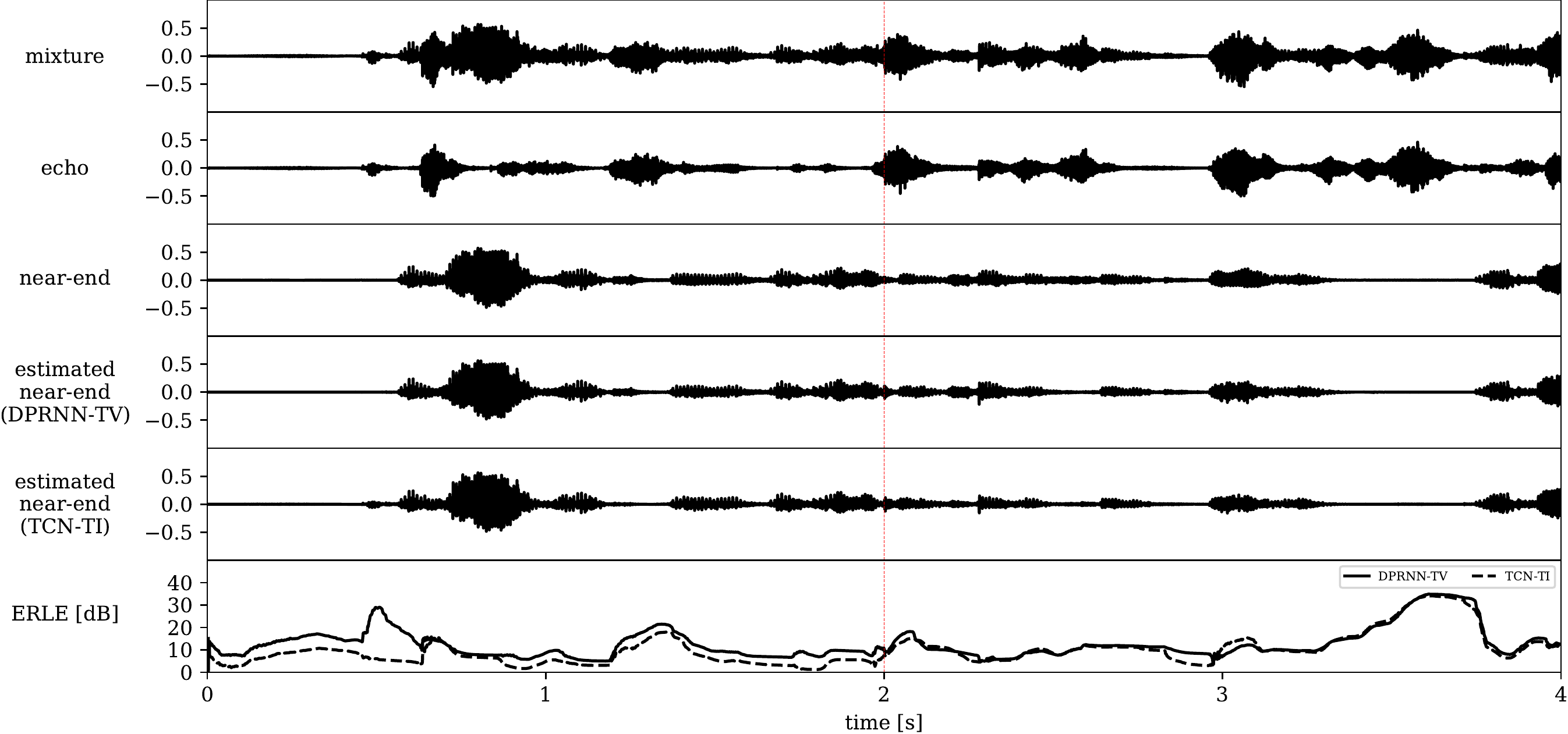}}
	\caption{Comparison between the proposed causal \ac{DPRNN}-TV model and the non-causal \ac{TCN}-TI model \cite{Delcroix2020} for a test example with 0 dB \ac{SIR}. The far-end signal comprises two non-overlapping speakers (2~s each). A sudden movement of the loudspeaker position occurs after two seconds (distance from the microphone is changed from 0.85 to 1.35 m).}
  \vspace{-0.7em}
	\label{fig:waveforms}%
\end{figure*}

Table~\ref{tab:resutls} shows the results for the different methods. We first observe high values for the NS and NN subsets which proves that learning target source characteristics from a reference snippet can be generalized to non-speech signals as well. Moreover, the proposed TV method significantly improves the extraction performance in all cases. This indeed suggests that the inclusion of the time-varying behavior of the learned embeddings is essential when a temporal correlation between the reference and the target signal exists. This is especially beneficial for the SS subset reaching an improvement of at least 4.5 dB. We also observe that both \ac{DPRNN}-TV and \ac{TCN}-TV models perform comparably well for the SS and SN subsets, whereas the former performs slightly better for the NS and NN subsets.

\begin{table}[t]
  \centering
  \resizebox{0.95\columnwidth}{!}{
  \begin{tabular}{@{}lc*{5}{S[table-format=2.2]}@{}}
\toprule
Method                             & Model Size  & Causal    & SS    & SN    & NS    & NN    \\ \midrule
\ac{TCN} - TI \cite{Delcroix2020} & 0.59M  &   $\times$   & 6.97 & 11.17 & 11.79 & 10.09 \\
\ac{TCN} - TV (proposed)          & 0.59M      &   $\times$   & \textbf{11.47} & \textbf{13.88} & 14.39 & 13.15 \\
\ac{DPRNN} - TI                   & 2.74M  &   $\times$   & 6.70 & 10.58 & 10.95 & 9.54 \\
\ac{DPRNN} - TV (proposed)        & 2.74M      &   $\times$   & 11.27 & 13.72 & \textbf{14.96} & \textbf{13.67} \\ \midrule
Zhang et al. \cite{Zhang2018}     & 8.95M &   $\times$   & 7.79 & 10.29 & 10.71 & 10.25 \\ \midrule
\ac{DPRNN} - TV (proposed)        & 2.10M   & $\text{\checkmark}$ & 10.47 & 12.87 & 13.75 & 12.22 \\ \bottomrule
\end{tabular}
}
\caption{Average \ac{SI-SDR} improvement [dB] results for the different methods. Evaluation is performed with respect to the near-end signal.}
\vspace{-1.0em}
\label{tab:resutls}
\end{table}


The proposed \ac{DPRNN}-\ac{TV} model significantly outperforms the \ac{AER} baseline \cite{Zhang2018} by at least 3.4 dB for the different subsets. This difference can be attributed to the way the reference signal is incorporated in the \ac{DNN}, where the authors in \cite{Zhang2018} used an early fusion mechanism, i.e., concatenation at the input space, which is shown to be inferior in \cite{vzmolikova2019speakerbeam}.

\begin{figure}[t]
\centerline{\includegraphics[width=.84\columnwidth]{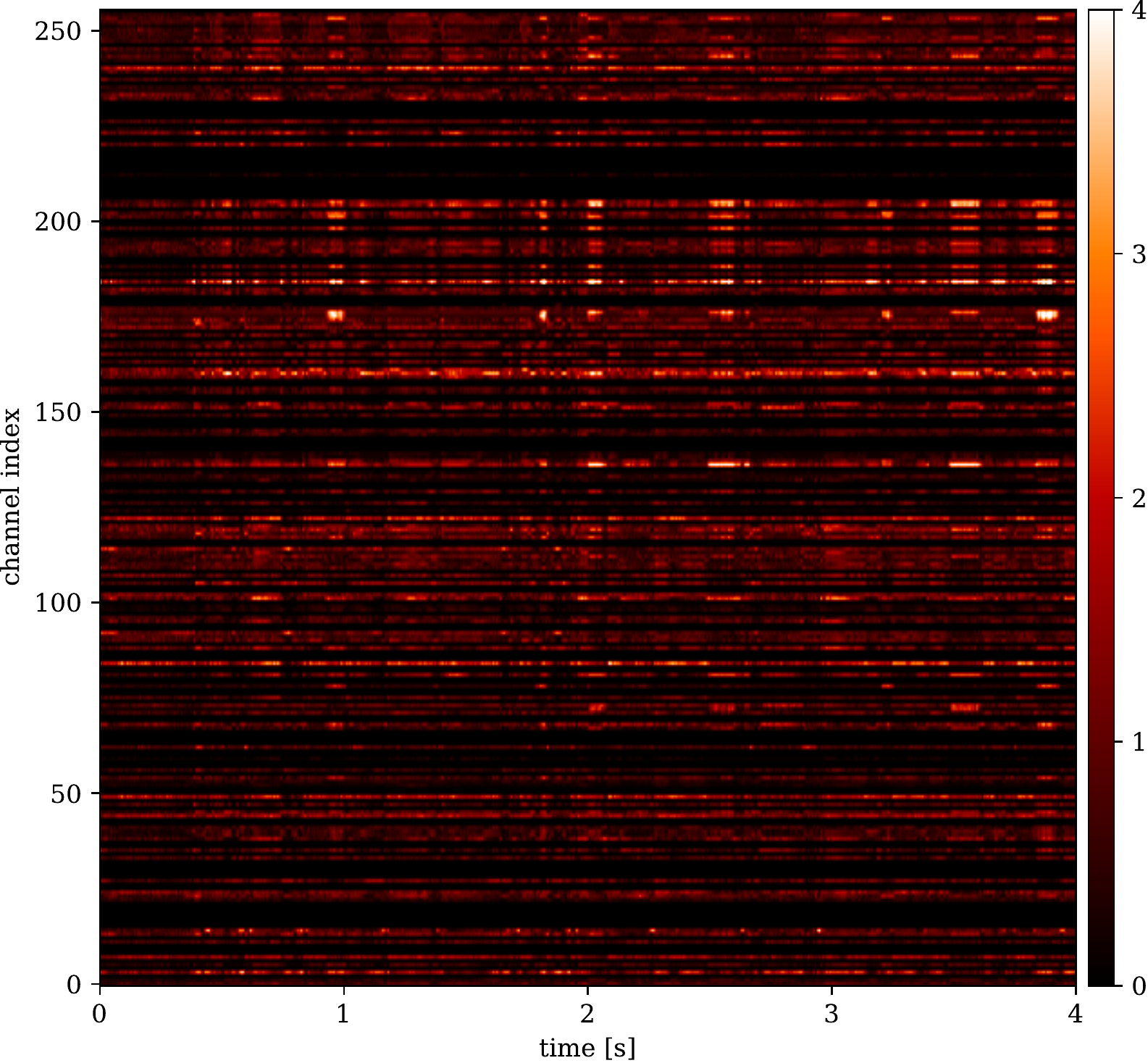}}
  \vspace{-0.2em}
  	\caption{Learned embeddings from the reference signal in the scenario shown in Figure~\ref{fig:waveforms}. To enhance visualizing the temporal behavior of the embeddings, the absolute value of the deviation from the mean (per channel) is depicted. }

  \vspace{-1.4em}
	\label{fig:embeddings}%
\end{figure}

The causal version of the proposed \ac{DPRNN}-TV model performs slightly worse than the non-causal model. However, it is interesting to observe that the causal model outperforms the non-causal \ac{TI} models in all subsets. We further evaluate the causal \ac{DPRNN}-TV model in an interesting scenario where the far-end signal comprises two different speakers one after the other. To make this scenario even more challenging, we simulate a sudden movement of the loudspeaker position, by using a different \ac{RIR} for the second speaker signal. This scenario was never included in the training stage. Such a scenario is challenging for the \ac{TI} method for two reasons. First, the time information of when each source starts is faded when averaging is applied, and secondly, the characteristics of the two sources can not be easily captured using one single embedding vector. Moreover, classical \ac{AER} methods that estimate the \ac{RIR} using adaptive filters can suffer from double-talk and, more severely, from sudden changes in the echo path during double-talk. Such points are mitigated by the proposed \ac{TV} method. Figure~\ref{fig:waveforms} illustrates the results of this scenario, in which a close resemblance between the estimated near-end signal using the proposed \ac{TV} method and the ground truth can be observed. In addition, it can be seen from the \ac{ERLE} \cite{enzner2014acoustic} plot that the proposed \ac{TV} method is not affected by the sudden change in the \ac{RIR} and the change in the speaker characteristics. Finally, Figure~\ref{fig:embeddings} shows the learned \ac{TV} discriminative embeddings for the reference signal in this scenario, where we clearly observe a time-varying behavior in some channels. However, it is interesting to see that a speaker change after 2~s produces a subtle difference in the learned embeddings.



\vspace{-0.8em}
\section{Conclusions}
In this paper, we proposed a \acf{TV} source discriminative model for informed source extraction that exploits the temporal dynamics of the reference signal. It has been shown that existing informed speaker extraction methods and the proposed method can be applied to non-speech signals as well. We applied the proposed \ac{TV} method in an \acl{AER} scenario and showed that it yields significantly better extraction results. Future work would investigate the impact of the temporal misalignment between the reference and the target signals on the performance.

\small
\bibliographystyle{ieeetr}
\bibliography{sapref}


\end{document}